\documentclass[aps,prb,twocolumn,groupedaddress]{revtex4}
\usepackage{amsmath}
\usepackage{amssymb}
\usepackage{graphicx}

\begin{document}

\title[Polarization-sensitive absorption of THz radiation in graphene]{Polarization-sensitive absorption of THz radiation 
by interacting electrons in chirally stacked
multilayer graphene}

\author{Maxim Trushin and John Schliemann}

\address{Institute for Theoretical Physics, University of Regensburg,
D-93040 Regensburg, Germany}

\begin{abstract}
We show that opacity of a clean multilayer graphene flake depends
on the helicity of the circular polarized electromagnetic radiation.
The effect can be understood in terms of the pseudospin selection rules for 
the interband optical transitions
in the presence of exchange electron-electron interactions which alter
the pseudospin texture in momentum space.
The interactions described within a semi-analytical 
Hartree--Fock approach
lead to the formation of the topologically different broken--symmetry states
characterized by Chern numbers and zero-field anomalous Hall conductivities.
\end{abstract}

\maketitle

\section{Introduction}

\begin{figure}
 \includegraphics[width=\columnwidth]{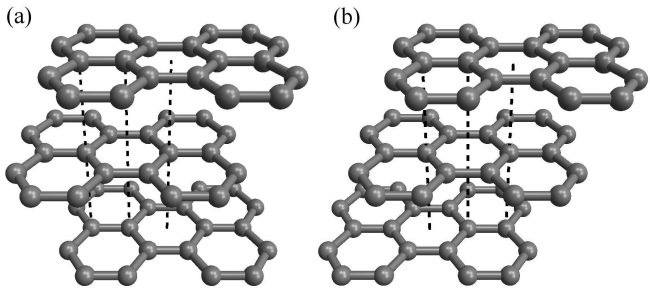}
\caption{(a) Bernal graphite represents graphene layers placed together in ...-AB-AB-AB-... stacking
when the two layers shifted with respect to each other by  $0.142\,\mathrm{nm}$
alternate in bulk.
(b) Rhombohedral graphite requires three non-equivalent graphene layers when
the two layers are shifted with respect to the first one
by  $0.142\,\mathrm{nm}$ and  $0.284\,\mathrm{nm}$
alternating in ...-ABC-ABC-ABC-... sequence. Dashed lines are shown for an eye guide.
Note that only ...-ABC-... stacking order results in the effective Hamiltonian (\ref{hN}) employed in this paper.}
\label{fig1ab}
\end{figure}

Multilayer graphene is a link between one atom thick 
carbon layers\cite{Nature2005novoselov}
 with the peculiar Dirac-like effective Hamiltonian for carriers\cite{RMP2009castroneto}
and graphite\cite{PR1947wallace} which 
can be seen as millions of graphene layers stacked together.
Graphene layers placed together do not lie exactly one on top of each other
but are shifted in such a way that only half of the carbon atoms
have a neighbor in another layer and the other half are projected
right into the middle of the graphene's ``honeycomb cell''.
If the third layer aligns with the first (and the $n+2$ layer with the $n$-th) then we arrive at
the more stable arrangement of graphene layers known as Bernal (or AB) stacking, see Fig.~\ref{fig1ab}a.
However, this is not the only possible configuration.
One can imagine an alternative stacking when the third layer aligns
neither the first nor the second but shifted with respect to both, see Fig.~\ref{fig1ab}b.
This arrangement is known as rhombohedral (or ABC) stacking and represents the main topic of this work.

The very first studies of Bernal and rhombohedral 
graphites\cite{PR1947wallace,PR1957mcclure,PRB1958slonczewski,CJP1958haering,Carbon1969mcclure}
relying on the tight binding model
have demonstrated the strong dependence of the band structure on the stacking order.
Later on the progress in numerical methods made it possible to refine 
the tight binding model outcomes using {\em ab initio} calculations \cite{PRB1970painter,PRB1988tomanek,PRB1991charlier}.
The seminal transport measurements on graphene\cite{Science2004novoselov} have inspired
recent investigations\cite{PRL2006latil,PRB2006partoens,PRB2007partoens,PRB2006guinea,PRB2010asvetisyan} of the band structure in a few layer graphene having
different stacking patterns. The band structure demonstrates a large variety of behavior
including a gapless spectrum, direct and indirect band gaps,
and energetic overlap of the conduction and valence bands
even though the number of layers has been limited by four.\cite{PRL2006latil}
Note that the band gap can be induced by an external electric field. \cite{PRB2010asvetisyan} 
The influence of an external magentic field (Landau levels) has also extensively been studied.
\cite{PRB2006guinea,PRB2011yuan2,NatPhys2011taychatanapat}
The electron-electron interactions have been taken into account in
Refs.~\cite{PRB2011yuan,PRB2010zhang2,PRL2011zhang,PRB2010nandkishore,2012lemonik},
also including the Zeeman term. \cite{2011zhang}
There are experimental indications that electron-electron interaction effects play an important
role in a few layer graphene where charge carriers may exhibit a variety of broken symmetry states.
\cite{Science2010weitz,NatPhys2011bao,NatNano2012velasco,Science2011mayorov}

\begin{figure}
 \includegraphics[width=\columnwidth]{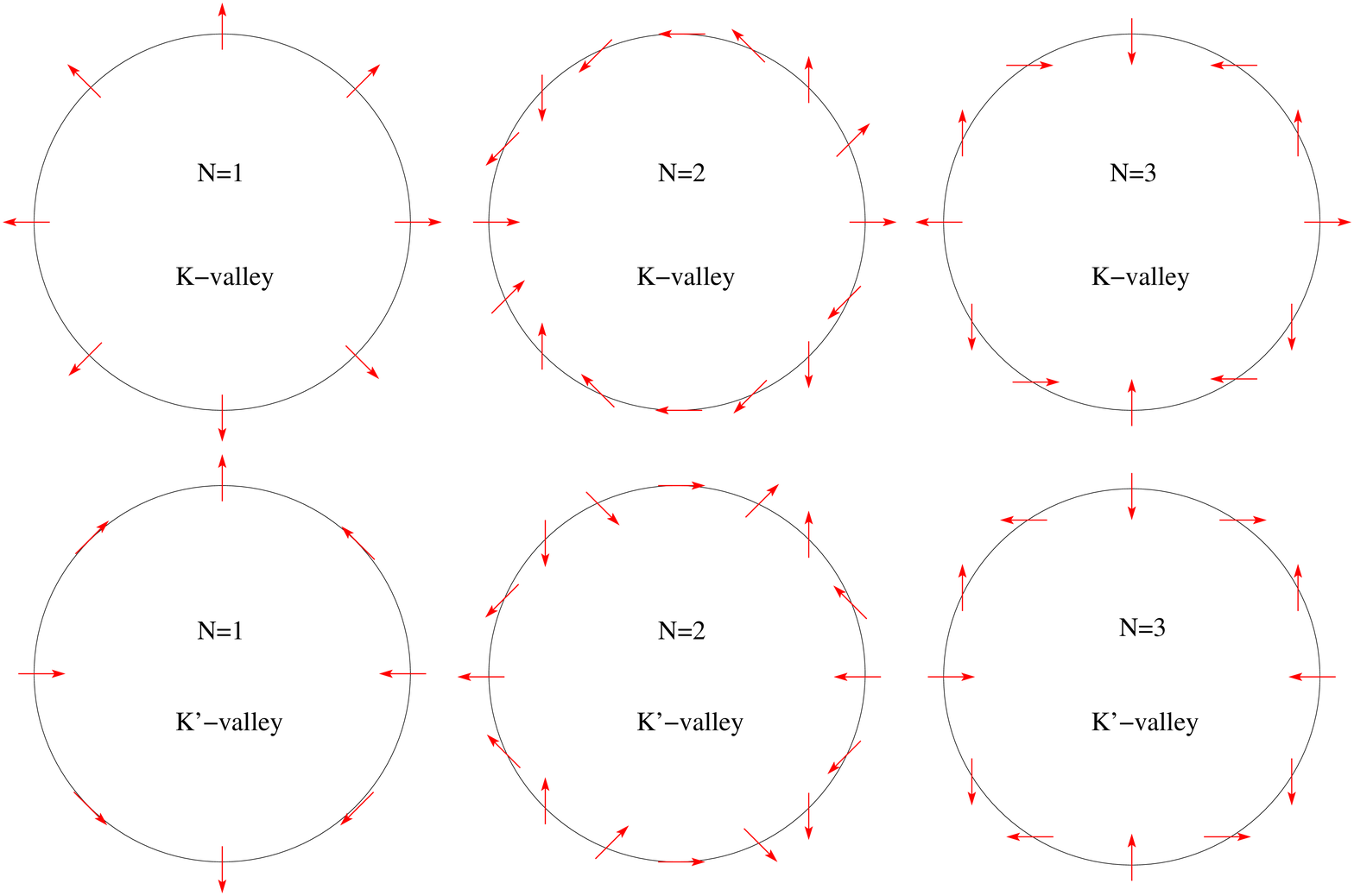}
\caption{Top view on the pseudospin texture for the conduction band in chirally stacked multilayer graphene
calculated from the non-interacting Hamiltonian (\ref{hN}). The pseudospin orientations in conduction
and valence bands for a given momentum are antiparallel.}
\label{fig1c}
\end{figure}

The brief literature overview given above illustrates how rich
the band structure of multilayer graphene (and the effects associated with) can be.
Note, however, that among all stacking possibilities, only the pure ABC arrangement
maintains the sublattice pseudospin chirality. \cite{PRB2010zhang}
In the simplest case of negligible interlayer asymmetries and trigonal warp
the simplified two-band ABC graphene model leads to the following Hamiltonian 
close to the neutrality point\cite{PRB2010zhang}
\begin{equation}
\label{hN}
H^\nu_0= \frac{(\hbar v_0)^N}{(-\gamma_1)^{N-1}}
\left(\begin{array}{cc}
0 
& (\nu k_x-i k_y)^N \\ 
(\nu k_x+i k_y)^N & 0
\end{array} \right),
\end{equation}
where $v_0\approx 10^6 \mathrm{ms^{-1}}$ is the group velocity for carriers in single layer graphene,
$\gamma_1$ is the hopping parameter, $\nu=\pm$ is the valley index, and $N$ is the number of layers.
As one can see from Eq.~(\ref{hN}),
the N-layer and (N+1)-layer graphene stacks differ,
apart from the density of states, by only the winding number
associated with the pseudospin orientation.
The pseudospin texture in the momentum space associated with Hamiltonian (\ref{hN}) and shown in Fig.~\ref{fig1c}
is the main topic of this work. 
In what follows we focus on the influence of electron-electron exchange interactions
on the pseudospin texture and its detection by optical means.

Note that the pseudospin lies in the $xy$-plane as long as its carrier remains in an eigenstate of $H^\nu_0$.
The exchange interactions can turn the pseudospin texture
to the out-of-plane phase with the out-of-plane angle depending on the absolute value of the particle momentum
\cite{PRL2011trushin,PRL2011zhang,PRB2011jung,PRB2008min}.
This is due to the huge negative contribution to the Hartree--Fock ground 
state energy from the valence band (i. e. ``antiparticle'' states) 
which cannot be neglected in graphene because of the zero gap
and conduction-valence band coupling via pseudospin.
The broken symmetry states in multilayer graphene with chiral stacking have been
recently studied by Zhang {\it et al} \cite{PRL2011zhang} using quite general arguments.
Spontaneous symmetry breaking can occur in presence of exchange electron-electron interactions \cite{PRB2008min}.
Here, we utilize a simplified model which includes exchange electron-electron interactions but, at the same time,
allows transparent half-analytical solution.
Having this solution at hand we focus on the manifestation of such broken symmetry states in optical absorption measurements.

Optical absorption via the direct interband optical transitions in monolayer graphene 
has been investigated in \cite{Science2008nair}
and shown to be equal to the universal value $\pi e^2/ \hbar c$.
In the presence of the electron-electron interactions
the interband absorption can be substantially reduced or enhanced
as compared to its universal value $\pi e^2/ \hbar c$ 
just by switching the helicity of the circularly polarized light \cite{PRL2011trushin}.
This effect is due to the peculiar pseudospin texture arising
from the interplay between pseudospin-momentum coupling
and exchange interactions.
To observe the pseudospin texture in chirally stacked multilayer graphene
by optical means the photon energy must be much smaller than the bottom of the lowest split-off bands
$\gamma_1=0.4\, \mathrm{eV}$. 
To give an example, a $\mathrm{CH_3OH}$ $20\,\mathrm{mW}$ laser \cite{2010krach} with wavelength
$118\,\mathrm{\mu m}$ (i.e. with the photon energy $10.5\,\mathrm{meV}$) safely satisfies this condition.

\section{Model}

We start from the Coulomb exchange Hamiltonian for chiral carriers which is given by
\begin{equation}
\label{ex}
H^\nu_\mathrm{exch}(\mathbf{k})=-\sum\limits_{\kappa'}\int\frac{d^2k'}{4\pi^2}
U_{|\mathbf{k}-\mathbf{k}'|} |\chi^\nu_{ k'\kappa'}\rangle\langle\chi^\nu_{ k'\kappa'}|
\end{equation}
with $U_{|\mathbf{k}-\mathbf{k}'|}=2\pi e^2/\varepsilon|\mathbf{k}-\mathbf{k}'|$  
and $\kappa'=\pm$ being the band index with $\kappa=+$ for the conduction band.
In order to consider the exchange Hamiltonian for any $N$ on the equal footing
we assume strictly two-dimensional Fourier transform for Coulomb potential.
This is in contrast to \cite{PRB2008min} where the interlayer distance  $d=0.335\,\mathrm{nm}$
has been included in the screening multiplier $\exp(-|\mathbf{k}-\mathbf{k}'|d)$.
Since the wave vector difference $|\mathbf{k}-\mathbf{k}'|$ can not be larger than the momentum cut-off 
of the order of $10^7\,\mathrm{cm}^{-1}$ employed in our model (see below)
the screening multiplier is always of the order of $1$ and can be disregarded here.
The intervalley overlap is also assumed to be negligible, and the eigenstates of
$H^\nu=H^\nu_0 +H^\nu_\mathrm{exch}$ can be formulated as
$\Psi_{\mathbf{k}\kappa}^\nu(\mathbf{r})={\mathrm e}^{i\mathbf{kr}}|\chi_{ k \kappa}^\nu\rangle$
with spinors 
$|\chi_{k+}^\nu\rangle=(\cos\frac{\vartheta_k}{2}, \nu\sin\frac{\vartheta_k}{2}\mathrm{e}^{\nu N i\varphi})^T$,
$|\chi_{k-}^\nu\rangle=(\sin\frac{\vartheta_k}{2}, -\nu\cos\frac{\vartheta_k}{2}\mathrm{e}^{\nu N i\varphi})^T$,
and $\tan\varphi=k_y/k_x$, where a non-zero out-of-plane pseudospin component 
corresponds to $\vartheta_k\neq \pi/2$.
To diagonalize $H^\nu$ the following $\nu$-independent equation for $\vartheta_k$ 
must be satisfied \cite{PRB2008juri,PRL2011trushin}
\begin{eqnarray}
\nonumber &&
\frac{(\hbar v_0 k)^N}{(-\gamma_1)^{N-1}}\cos\vartheta_k+\sum\limits_{\kappa'}
\int\frac{d^2k'}{8\pi^2}\kappa'U_{|\mathbf{k}-\mathbf{k}'|}
\left[\cos\vartheta_{k'}\sin\vartheta_k - \right. \\
&& \left. \sin\vartheta_{k'}\cos\vartheta_k\cos(N \varphi'-N \varphi) \right] =0.
\label{thetak}
\end{eqnarray}
Here the integration goes over the occupied states.
Note that the conduction and valence states are entangled, 
and the latter cannot be disregarded even at positive Fermi energies assumed below.
Thus, in order to evaluate the integrals in Eq.~(\ref{thetak}) a momentum
cut-off $\Lambda$ is necessary. Most natural choice $\Lambda=\gamma_1/\hbar v_0$ 
corresponds to the energy scale $\gamma_1$ at which the split-off bands of
bilayer graphene become relevant and our two-band model no longer applies.
Substituting $x=k/\Lambda$ we arrive at
\begin{eqnarray}
\label{x} &&
\frac{4\pi}{\alpha^*}x^N\cos\vartheta_k  =\\
\nonumber &&  \int\limits_0^{2\pi} d\varphi'
\int\limits_{k_F/\Lambda}^{1}dx'x'
\frac{\cos\vartheta_{k'}\sin\vartheta_k 
- \sin\vartheta_{k'}\cos\vartheta_k\cos N\varphi'}
{\sqrt{x^2+x'^2-2xx'\cos\varphi'}},
\end{eqnarray}
where $\alpha^*=e^2/(\varepsilon \hbar v_0)$.
The momentum cut-off is assumed to be much larger than the Fermi momentum $k_F$,
and, therefore, we can set the lower integral limit to zero.
In this case our outcomes do not depend on the value of $\Lambda$.

\begin{figure}
\includegraphics[width=0.49\columnwidth]{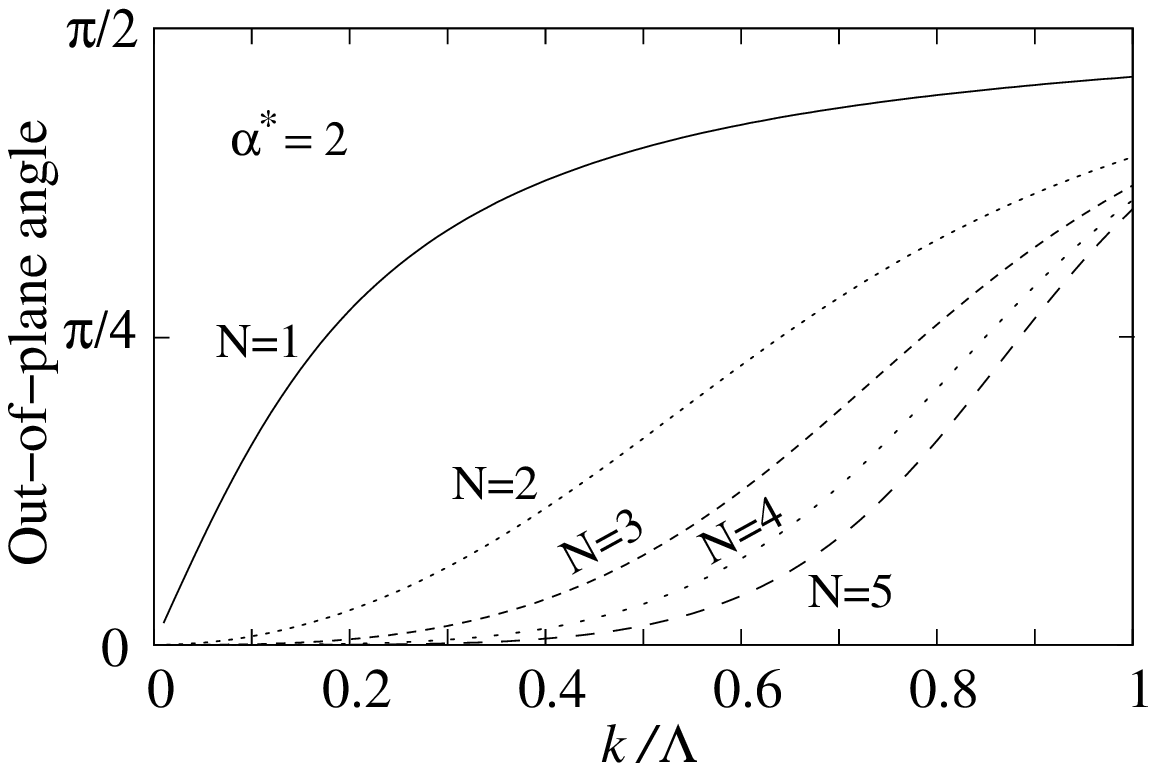}
\includegraphics[width=0.49\columnwidth]{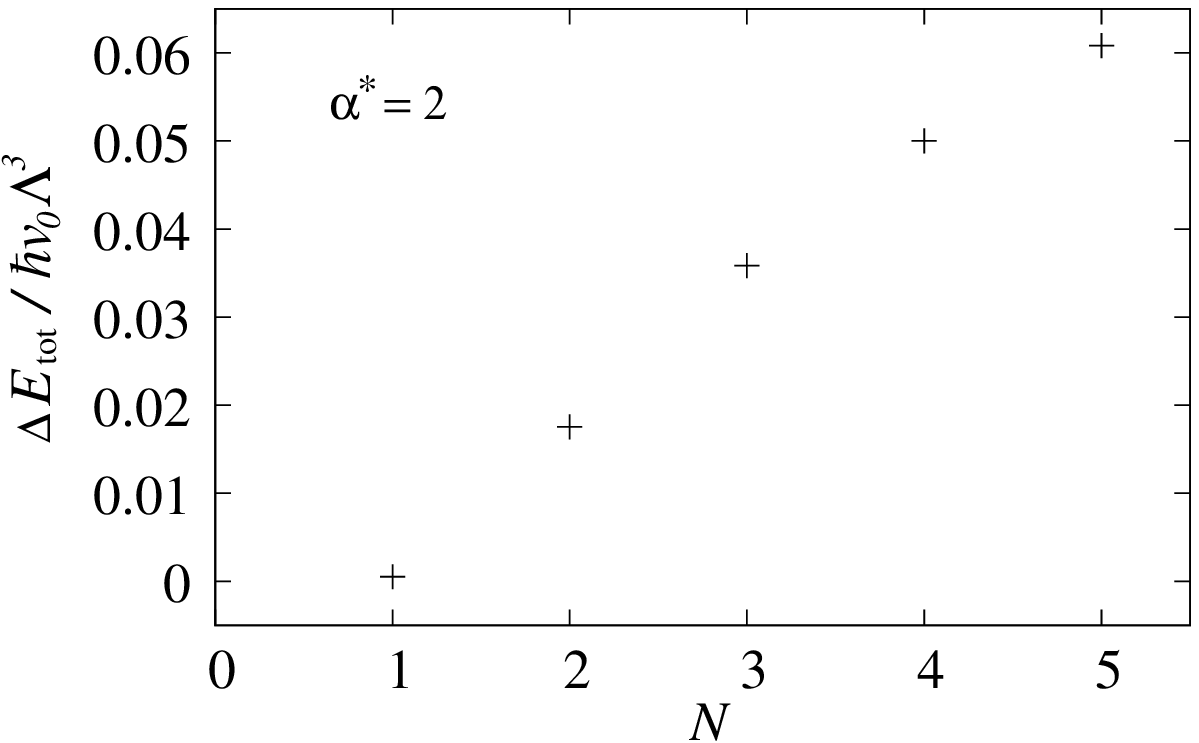}
\caption{Left panel: The pseudospin out-of-plane angle $\vartheta(k)$ 
for substrate-free chirally stacked $N$-layer graphene numerically calculated from Eq.~(\ref{x}).
Right panel: The total ground state energy difference (\ref{Etot}) 
between the in-plane and out-of-plane phases for different $N$.
Increasing $N$ makes the out-of-plane phase more preferable.
The $\vartheta(k)$ curve for $N=1$ differs from the one given in \cite{PRL2011trushin}
since we have improved the precision of our calculations here.
Note, that the Coulomb interactions are completely unscreened here.
The screening is taken into account in Fig.~\ref{fig3ab} by introducing
the effective dielectric constant $\varepsilon$ which relates to  
 $\alpha^*$ as $\alpha^*=e^2/(\varepsilon \hbar v_0)$.}
\label{fig2ab}
\end{figure}

\begin{figure}
\includegraphics[width=0.49\columnwidth]{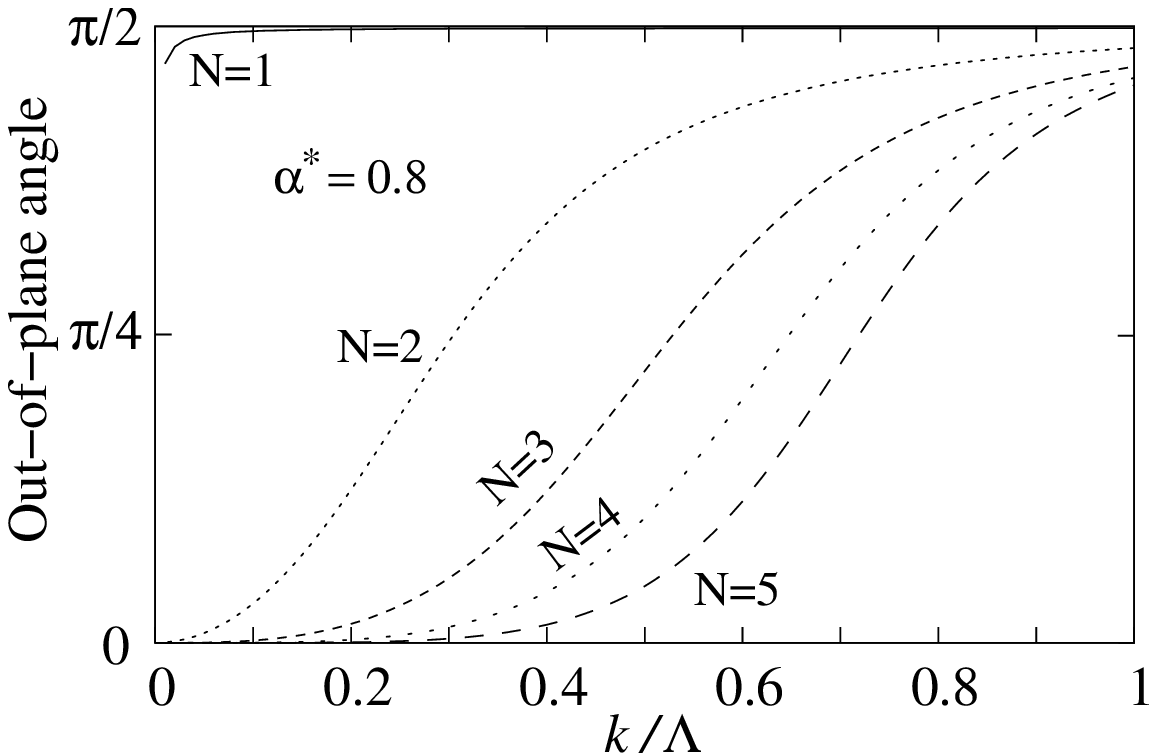}
\includegraphics[width=0.49\columnwidth]{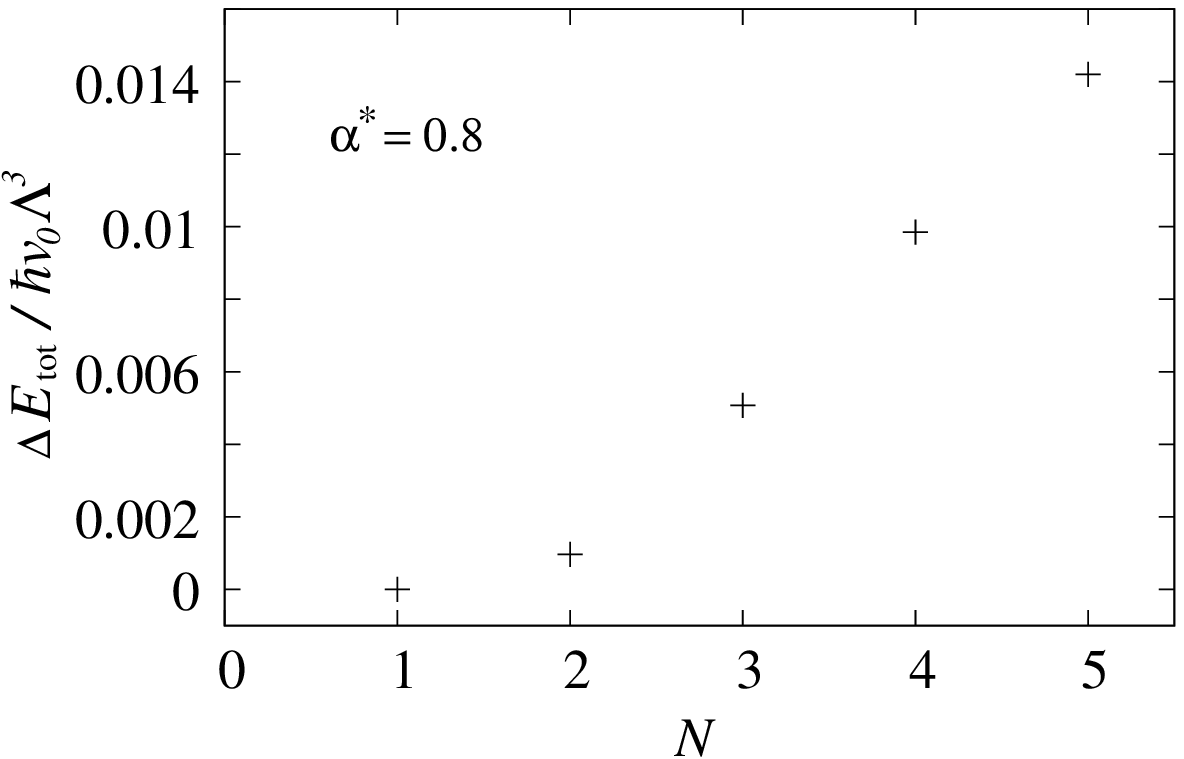}
\caption{Left panel: The pseudospin out-of-plane angle $\vartheta(k)$ 
for chirally stacked $N$-layer graphene on $\mathrm{SiO}_2$ numerically calculated from Eq.~(\ref{x}).
Right panel: The total ground state energy difference (\ref{Etot}) 
between the in-plane and out-of-plane phases for different $N$.
The difference is much larger in the case of substrate-free graphene,
see Fig.~\ref{fig2ab}.}
\label{fig3ab}
\end{figure}

Besides a trivial solution with $\vartheta_0=\pi/2$ independent of $k$, there are
non-trivial ones $\vartheta_1=\vartheta(k)$ and $\vartheta_2=\pi-\vartheta(k)$
with $\vartheta(k)$ shown in Figs.~\ref{fig2ab}a and \ref{fig3ab}a for different $N$
and $\alpha^*$. The solutions $\vartheta_0$ and $\vartheta_{1,2}$ represent to two phases with different total ground 
state energies $E_\mathrm{tot}^\mathrm{in}$ ($E_\mathrm{tot}^\mathrm{out}$) for the in-plane 
(out-of-plane) pseudospin phase. The difference 
$\Delta E_\mathrm{tot} =E_\mathrm{tot}^\mathrm{in}-E_\mathrm{tot}^\mathrm{out}$
per volume is given by 
\begin{eqnarray}
\nonumber &&
\frac{\Delta E_\mathrm{tot}}{\hbar v_0 \Lambda^3} 
= -\frac{g_s g_\nu}{ 2\pi}\int\limits_0^1 dx' x'^{N+1}(1-\sin\vartheta_{k'}) \\
\nonumber &&
-\alpha^* \frac{g_s g_\nu}{32\pi^3}\int\limits_0^{2\pi} d\varphi \int\limits_0^{2\pi} d\varphi'
\int\limits_0^{1}dx\int\limits_0^{1}dx' \\
\nonumber && \times xx'\frac{(1-\sin\vartheta_{k'}\sin\vartheta_k)
\cos(N\varphi'-N\varphi)-\cos\vartheta_{k'}\cos\vartheta_k}
{\sqrt{x^2+x'^2-2xx'\cos(\varphi-\varphi')}},\\
\label{Etot}
\end{eqnarray}
The ground state energy is the same for {\em both} valleys and spins,
and, therefore, Eq.~(\ref{Etot}) contain $g_s=2$ and $g_v=2$ for spin and valley degeneracy respectively.
$\Delta E_\mathrm{tot}$ has been evaluated numerically and the resulting 
$\Delta E_\mathrm{tot}$ vs. $N$ dependency is shown in Figs.~\ref{fig2ab}b and \ref{fig3ab}b
for suspended and $\mathrm{SiO}_2$-placed graphene respectively.
One can see that strong electron-electron interactions with $\alpha^*=2$ definitely make the out-of-plane
phase energetically preferable for $N>1$. 
For graphene placed on $\mathrm{SiO}_2$ substrate the pseudospin out-of-plane phase is energetically preferable only for $N\geq 3$.
Note that the estimates of $\alpha^*$ for clean monoatomic graphene flake vary from $2$ 
(Ref.~\cite{PRL2008jang}) to $2.8$ (Ref.~\cite{PRB2005peres}) and, therefore,
even monolayer graphene may get to the pseudospin out-of-plane phase.

Note that it is possible to choose either the same or opposite solutions for
two valleys. The former choice breaks the parity invariance whereas
the latter one does so with the time reversal symmetry.\cite{PRL2011trushin}
In what follows we consider possible manifestations of these solutions
in the optical absorption measurements on multilayer graphene.

\section{Optical absorption}

We assume that an electromagnetic wave is propagating along $z$ axis
perpendicular to graphene plane.
It can be described by the electric field $\mathbf{E}=\mathbf{E}_0 \exp(i k_z z -i\omega t)$,
where $k_z$ is the wave vector, and $\omega$ is the radiation frequency.
For our purpose it is more convenient to characterize the electromagnetic wave
by its vector potential $\mathbf{A}=\mathbf{A}_0 \exp(i k_z z -i\omega t)$,
where $\mathbf{A}_0 = ic\mathbf{E}/\omega$. The relation between $\mathbf{E}$
and $\mathbf{A}$ is unambiguous since $\mathbf{A}$ is always in $xy$ plane 
resulting in a gauge with $\mathrm{div} \mathbf{A}=0$.
The interaction Hamiltonian can be then written in the lowest order in  $\mathbf{A}$ as
\begin{eqnarray}
\label{int}
&& H^\nu_\mathrm{int}= \frac{(\nu \hbar v_0 k)^N }{k(-\gamma_1)^{N-1}} \frac{eN}{\hbar c} \\
\nonumber &&
\times \left(\begin{array}{cc}
0 
& {\mathrm e}^{-\nu i \varphi (N-1)} (A_x- i \nu A_y) \\ 
{\mathrm e}^{\nu i \varphi (N-1)} (A_x+ i \nu A_y) & 0
\end{array} \right).
\end{eqnarray}
The total absorption $P$ can be calculated as a ratio between 
the total electromagnetic power absorbed
by graphene per unit square  $W_a=\hbar\omega\sum_\nu\int\frac{d^2k}{4\pi^2}w(\mathbf{k}+,\mathbf{k}-)$
and the incident energy flux  $W_i=\omega^2 A^2/4\pi c$. 
The probability to excite an electron from the valence band to an 
unoccupied state in the conduction band $w(\mathbf{k}+,\mathbf{k}-)$ can be calculated using Fermi's golden-rule
$$
w(\mathbf{k}+,\mathbf{k}-)=\frac{2\pi}{\hbar}|\langle \mathbf{k}+| H^\nu_\mathrm{int}|\mathbf{k}-\rangle |^2
\delta(E_{k+} - E_{k -} -\hbar\omega).
$$
Because of normal incidence there is zero momentum transfer from photons to electrons.
What is important is that the inter-band transition matrix elements of $H^\nu_\mathrm{int}$ turn out to be sensitive to
the light polarization and pseudospin orientations in the initial and final states.
In particular, for linear polarization with $A_x=A\cos\varphi_\mathrm{pol}$
and $A_y=A\sin\varphi_\mathrm{pol}$ the probability to excite an electron with a given momentum reads
\begin{eqnarray}
\nonumber && w(\mathbf{k}+,\mathbf{k}-)|_{\nu=+}
= \frac{\pi}{\hbar}\left(\frac{eN}{\hbar c}A\right)^{2} 
\frac{(\hbar v_0)^{2N} k^{2N-2}}{(-\gamma_1)^{2N-2}} \\
\nonumber &&
\times \delta(E_{k+} - E_{k -} -\hbar\omega) \\
&&\times \left[ 1+\cos^2\vartheta_k -\sin^2\vartheta_k \cos(2N\varphi-2N\varphi_\mathrm{pol}) \right].
\label{probabil1}
\end{eqnarray}
Note that $w(\mathbf{k}+,\mathbf{k}-)$ does not depend on the valley index $\nu$. 
The total optical absorption $P$ is not sensitive to the particular orientation of the polarization plane since
the $\varphi_\mathrm{pol}$-dependent term is integrated out in this case.
In contrast, if we assume a circular polarization fulfilling $A_x=\pm iA/\sqrt{2}$, $A_y=A/\sqrt{2}$
then the interband transition probability for K-valley can be written as
\begin{eqnarray}
\nonumber && w(\mathbf{k}+,\mathbf{k}-)|_{\nu=+}
= \frac{4\pi}{\hbar}\left(\frac{eN}{\hbar c}A\right)^{2} 
\frac{(\hbar v_0)^{2N} k^{2N-2}}{(-\gamma_1)^{2N-2}} \\
&& \times
\left\{ \begin{array}{c}
\sin^4\frac{\vartheta_k}{2} \\ 
\cos^4\frac{\vartheta_k}{2}
\end{array} \right\} 
 \delta(E_{k+} - E_{k -} -\hbar\omega).
\label{probabil2}
\end{eqnarray}
Here, the multipliers $\sin^4(\vartheta_k/2)$ and $\cos^4(\vartheta_k/2)$ are
for two opposite helicities of light, and for K'-valley they are interchanged.
If the out-of-plane pseudospin polarization is chosen to be opposite in two valleys
then this helicity dependence survives the integration over momentum and
summation over valley index resulting in the helicity-sensitive total absorption which reads
\begin{eqnarray}
\label{P}
\nonumber && P=\frac{16 N^2}{\hbar \omega} \frac{\pi e^2}{\hbar c}  \frac{(\hbar v_0)^{2N}}{(-\gamma_1)^{2N-2}}
\int\limits_0^\Lambda dk k^{2N-1}
\left\{ \begin{array}{c}
\sin^4\frac{\vartheta_k}{2} \\ 
\cos^4\frac{\vartheta_k}{2}
\end{array} \right\} \\
 &&  \times \delta(E_{k+} - E_{k -} -\hbar\omega).
\end{eqnarray}
The absorption strongly depends on helicity
as long as the radiation frequency is much smaller than the band split-off energy.
This regime corresponds to the excitation of electrons with comparatively small momenta,
where the angle $\vartheta_k$ is close to zero, see Fig.~\ref{fig2ab}b and Fig.~\ref{fig3ab}b.
In the in-plane phase with $\vartheta_k=\pi/2$ the total absorption does not depend on
light polarization, and in the non-interacting limit it equals to 
\begin{equation}
P=N \frac{\pi e^2}{\hbar c}
\label{noint}
\end{equation}
At $N=1$ it acquires the universal value $\frac{\pi e^2}{\hbar c}$, as expected\cite{Science2008nair}.

The experiment proposed here is similar to the one discussed recently in 
Ref.~\cite{PRL2011nandkishore} suggesting to detect 
broken symmetry states in graphene placed on substrate
by polar Kerr rotation measurements, i.e. by analyzing the elliptical
polarization of reflected radiation. In the present paper, however, we 
propose to reach the same goal by comparing the opacity for two orthogonal
light polarizations, which appears to be the simpler strategy.

Alternatively, one can characterize broken--symmetry states in terms of the 
topological Chern numbers \cite{PRL1982thouless} given by 
\begin{equation}
 \label{chern1}
n^\nu_\pm=\frac{1}{2\pi}\int d^2 k \nabla_k \times \mathbf{A}^\nu_\pm
\end{equation}
with $\mathbf{A}^\nu_\pm=i\langle\chi_{k\pm}^\nu|\nabla_k|\chi_{k\pm}^\nu\rangle$.
This approach is routinely used in the theory of topological 
insulators \cite{RMP2010hasan} and originated from the description of
quantized Hall conductances.

A non-trivial band topology has also been found in graphene 
\cite{PRL2005kane,SSC2011prada}
where the band gap was opened via spin-orbit coupling characterized just by a 
constant (i.e. wave vector independent) mass term
rather than by exchange interactions described by a more complicated 
Hamiltonian (\ref{ex}).

Note, that in our case here $\nabla_k \times \mathbf{A}^\nu_\pm$ contains a 
term proportional to $\delta(\mathbf{k})$
which occurs due to the singularity  in $\mathbf{A}^\nu_\pm \propto k^{-1}$
but does not contribute to $n^\nu_\pm$ as long as $\vartheta_k=0$ at $k\to 0$.
Thus, for the conduction band electrons we have
\begin{equation}
 \label{chern2}
n^\nu_+=-\frac{\nu N}{2}\int\limits_{\vartheta(0)}^{{\vartheta(\Lambda)}} d \vartheta_k \sin\vartheta_k.
\end{equation}
Assuming the out-of-plane pseudospin polarization to be opposite in two valleys
it follows that the total Chern number for conduction electrons just equals 
$N$. On the other hand, the total Chern number is zero as long as the 
out-of-plane pseudospin pseudospin
component is same in both valleys. Thus, the two out-of-plane solutions
are {\em topologically} different even though they correspond to the band gap
of the same size. The topologically non-trivial case with $n\neq 0$
corresponds to broken time reversal symmetry leading to the 
existence of a zero-field Hall current \cite{PRL2011trushin}. 

The only question is whether the time reversal broken states really occur in clean graphene samples.
More in depth analysis performed by Nandkishore and Levitov \cite{PRL2011nandkishore,PRB2010nandkishore}
suggests that this is the case at least for bilayer graphene.
On the other hand Jung {\it et al.}\cite{PRB2011jung} demonstrate that the intervalley exchange coupling 
favors parallel pseudospin polarization in opposite valleys breaking the parity invariance.
In this case the total absorption does not depend on the radiation helicity but
the two valleys are  occupied  differently by the photoexcited carriers
which might be useful effect for {\em valleytronics} \cite{PRL2007xiao}.
Note that the strain effects also may change the band structure topology in bilayer \cite{PRB2011mucha}
and probably multilayer graphene. This might be an issue in suspended samples
where mechanical deformations can occur easily.
There is, however, an experimental evidence \cite{Science2011mayorov} of the fact that it is the electron-electron interaction
rather than the strain that is responsible for the band structure reconstruction.

\section{Conclusion}

We have demonstrated that the polarization-sensitive optical
absorption predicted in \cite{PRL2011trushin} for single layer graphene
can be found in chirally stacked carbon multilayers within a broader range of parameters.
This is due to the enhancement of the interaction effects
in multilayer graphene with larger number of layers.
The conditions necessary for the observation
of the polarization-dependent optical absorption can
be summarized as follows. (i) Graphene samples must be 
prepared as cleaner as possible.
It would be better to utilize suspended samples in order
to isolate the interacting electrons from the environment
with large relative permittivity.
(ii) The exchange interactions must break the time reversal
invariance rather than the parity. This situation corresponds
to the topologically non-trivial state with the Chern number equal to $N$.
(iii) Finally, the low frequency radiation is necessary
to excite the electrons with smaller momenta having larger
out-of-plane pseudospin components and
to preclude the influence of split-off bands.

In addition, we would like to mention the possible existence of
a zero-field Hall current in the slightly doped graphene samples where
the time reversal invariance is broken as described above.

\section*{Acknowledgements}

We would like to thank Tobias Stauber, Jeil Jung, and Fan Zhang for stimulating discussions.
This work was supported by DFG via GRK~1570.

\section*{References}

\bibliography{graphene.bib,optical.bib}

\end{document}